# New results of the search for hidden photons by means of a multicathode counter.

A.V.Kopylov, I.V.Orekhov, V.V.Petukhov

Institute for Nuclear Research of RAS, 117312 Moscow, Russian Federation,

Prospect of 6oth Anniversary of October Revolution 7A

Abstract

New upper limit on a mixing parameter for hidden photons with a mass from 5 eV till 10 keV has been obtained from the results of measurements during 78 days in two configurations $R_1$ and $R_2$ of a multicathode counter. For a region of a maximal sensitivity from 10 eV till 30 eV the upper limit obtained is less than $4 \times 10^{-11}$. The measurements have been performed at three temperatures: 26°C, 31°C and 36°C. A positive effect for the spontaneous emission of single electrons has been obtained at the level of more than 7σ. A falling tendency of a temperature dependence of the spontaneous emission rate indicates that the effect of thermal emission from a copper cathode can be neglected.

Hidden photons have been first suggested by L.B.Okun' [1] as a modification of electrodynamics. In a model when the hidden photons of a certain mass constitute a cold dark matter (CDM) one should register in experiment the effect from a kinetic mixing of a hidden photon with the ordinary electromagnetic field. In this case a field of photons is of a spatially constant mode $\mathbf{k} = 0$ oscillating with frequency $\omega = m_{\gamma'}$, where $m_{\gamma'}$ – mass of a hidden photon, χ – a mixing parameter.

$$\left.\begin{pmatrix} \mathbf{A} \\ \mathbf{X} \end{pmatrix}\right|_{DM} = \mathbf{X}_{DM} \begin{pmatrix} -\chi \\ 1 \end{pmatrix} \exp(-i\omega t) \qquad (1)$$

Here **X** – field of the hidden photons, **A** – field of the real photons. To register the hidden photons with energy greater than a work function of a material of a cathode a multicathode counter has been developed by us [3]. The signal from hidden photons is measured in this technique by difference of count rates ($R_1 - R_2$) of single electrons in the configurations 1 and 2. In the first case, as one see at Fig.1, the cathode C3 is under potential which enables a drift of electrons, emitted from a cathode, in the direction to central counter with the cathode C1. In the second case the cathode C3 is under potential which reject electrons from moving in this direction. As a result the magnitude of a signal is determined here as

$$R_{MCC} = (R_1 - R_2)/\varepsilon \qquad (2)$$

Here : ε – the efficiency of the counting of single electrons determined from calibration of the counter by UV photons of a mercury lamp.

A design of the counter and the description of a method have been described in details in [3, 4]. The most essential features are that a central counter with a cathode C1 has a high (>$10^5$) coefficient of gas amplification, what enabled to register single electron events. The volume between cathodes C1 and C3 has been used to transport electrons to a central counter. Due to this volume it becomes possible to increase substantially a cathode's surface which is 2000 $cm^2$ what guarantees high sensitivity of this detector. A charge sensitive preamplifier Zaryad produced in Russia has been used which had a sensitivity 0.5 V/PC and noise <100 μV. A noise has been determined mainly by electronic circuit and by 8-bit ADC board

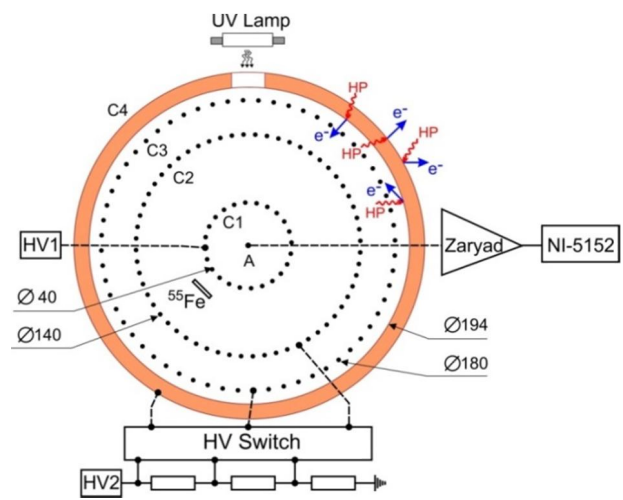

Figure.1. A simplified scheme of a multicathode counter.

In this experiment 8-bit Digitizer NI PCI-5152 working in a range of ± 50 мВ. The data has been collected by frames, each frame contained 2·$10^6$ points with intervals 100 nsec. This massive of data has been written down first in the internal memory of ADC board. The collection of data has been stopped at this moment, then data heve been written to a hard disk then the collection of the data resumed. During each 24 hours two files have been obtained, one during a night and one during a day, each one contains 5000 frames of the total capacity of about 200 GB. The total volume of the information collected is limited mainly by the volume of internal memory of ADC, the velocity of a data transform through a bus of PCI and by a performance of PC itself.

The data treatment has been conducted in off line. The frames with excessive noise are discarded. A useful signal is considered to be when amplitude is between 3 and 30 mV with a short front edge corresponding to a time of a drift of positive ions to cathode, with a slow tail corresponding to a time of a restoration of a baseline of a charge sensitive preamplifier with undistorted baseline. Following the procedure of data treatment we obtain two experimental points for each day of measurements. In a paper [4] we have published an upper limit obtained from the data collected during 28 days. From these measurements the count rate $R_{MCC}$ = - 0.06 ± 0.36 Hz has been obtained. From here we have got the limit $R_{MCC}$ < 0.66 Hz for 2σ (95%) confidence level. The upper limit for mixing parameter χ for a mass of hidden photon from 5 eV till 10 keV is presented by dashed line at Fig. 2. A ragged line corresponds to quantum efficiencies taken for atomically clean copper, smooth line – to a copper routinely cleaned when the impurities in copper produce a smoothing effect. The data do not match at certain energies because the quantum efficiencies taken from different experiments differ. By calculation the upper limits for χ we have used the expression [3, 4]

$$\chi_{sens} = 2.9 \times 10^{-12} \left( \frac{R_{MCC}}{\eta \, 1 \, \text{Hz}} \right)^{1/2} \left( \frac{m_{\gamma'}}{1 \, \text{eV}} \right)^{1/2} \left( \frac{0.3 \, GeV/cm^3}{\rho_{CDM}} \right)^{1/2} \left( \frac{1 \, \text{m}^2}{A_{MCC}} \right)^{1/2} \left( \frac{\sqrt{2/3}}{\alpha} \right) \quad (3)$$

Here $A_{MCC}$ – the surface of the counter, α – a parameter of anisotropy of the flux of hidden photons, for isotropic case α = $\sqrt{2/3}$, , η – quantum efficiency, data taken from [5 – 8].

In 2016 we have performed a new run of measurements during 78 days. A massive of the data of about 15 TB has been analyzed. The calibration and measurements have been conducted by the same technique described in [3, 4]. An optimal mode of the detector operation has been used upon the obtained results of technical study. The measurements have been conducted at temperatures: 26°C, 31°C and 36°C. The analyses of the data produced the following rates in Hz per cm² of the surface of a cathode: $r_{MCC}$(26°C) = (0.98 ± 0.22)·10⁻⁴ , $r_{MCC}$(31°C) = (0.75 ± 0.15)·10⁻⁴ , $r_{MCC}$(36°C) = (0.69 ± 0.23)·10⁻⁴. The measured rate decreased with temperature what proves that the effect of thermal emission from a copper cathode can be neglected. The average value obtained from these three values is $r_{MCC}$ = (0.79 ± 0.11)·10⁻⁴ Hz/cm². One can see that the measured effect is above zero at more than 7 standard deviations. From here the upper limit has been obtained for 2σ( 95% ) confidence level which is depicted by solid line at Fig 2. One should consider these results as very preliminary. To test these results we are planning to perform measurements at lower temperature and also to use a new detector of a developed design. The value of this result is that it was obtained in direct measurements of the rate of single electrons from the surface of a metallic cathode. The limits obtained using detectors with volume as an active medium are certainly more rigid [9, 10]. But, as it has been understood in 30tees of last century, the processes on a surface are very different by physics from the ones in a volume and are very sensitive to the details of theoretical models [11]. This difference can lead to the different results, so it is very important to clarify details by experiment.

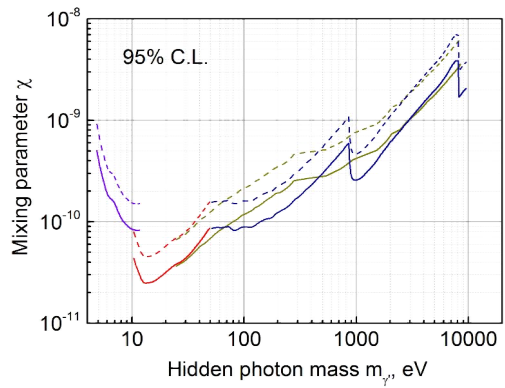

Figure.2 The upper limit for mixing parameter χ obtained. A dashed line – the results of 2015 yr [3, 4], a solid line – the results of this work.